 \documentclass[final,5p,times,twocolumn,12pt,sort&compress]{elsarticle}
\usepackage{soul}
\usepackage{bm}
\usepackage{ulem}
\usepackage{lineno,hyperref}
\usepackage{txfonts}
\usepackage{graphicx}
\usepackage{dcolumn}% Align table columns on decimal point
\usepackage{array}
\usepackage{color}
\usepackage{multicol}
\usepackage{amssymb}
\usepackage{amsmath}
\usepackage{type1cm}
\usepackage{caption}
\usepackage{natbib}
\usepackage{lineno,hyperref}
\usepackage{comment}
\usepackage{longtable}
\usepackage{ctable}
\modulolinenumbers[5]

\onecolumn

\journal{JQSRT, prepared by using elsarticle.cls}
\begin{document}
\begin{frontmatter}
\title{Extended calculations of energy levels, radiative properties, $A_{J}$, $B_{J}$ hyperfine
interaction constants, and Land\'e $g_{J}$-factors for
nitrogen-like \mbox{Ge XXVI}}

\author[hb,mm]{K. Wang}
\author[fd]{C. Y. Zhang}
\author[mm]{P. J\"{o}nsson\corref{per}}
%\author[hb]{P. Yang}
%\author[hb]{L. Guan}
\author[fd]{R. Si}
\author[hb]{X. H. Zhao}
%\author[hbg]{X. Yang}
\author[nudt]{Z. B. Chen\corref{bin}}
\author[ch]{X. L. Guo\corref{guo}}
\author[fd]{C. Y. Chen}
\author[bj]{J. Yan}

\cortext[per]{per.jonsson@mah.se} 
\cortext[bin]{chenzb008@qq.com} 
\cortext[guo]{xlguo12@fudan.edu.cn}

\address[hb]{Hebei Key Lab of Optic-electronic Information and Materials, The College of Physics Science and Technology, Hebei University, Baoding 071002, China}
\address[mm]{Group for Materials Science and Applied Mathematics, Malm\"o University, SE-20506, Malm\"o, Sweden}
\address[fd]{Shanghai EBIT Lab, Institute of Modern Physics, Department of Nuclear Science and Technology, Fudan University, Shanghai 200433, China}
\address[nudt]{School of Science, Hunan University of Technology, Zhuzhou, 412007, China}
\address[ch]{Department of Radiotherapy, Shanghai Changhai Hospital, Shanghai 200433, China}
\address[bj]{Institute of Applied Physics and Computational Mathematics, Beijing 100088, China}

%\address[pk]{Center for Applied Physics and Technology, Peking University, Beijing 100871, China}
%\address[jt]{Collaborative Innovation Center of IFSA (CICIFSA), Shanghai Jiao Tong University, Shanghai 200240, China}
%\address[hbg]{The Third Institute of Surveying and Mapping of Hebei Province, Hebei Bureau of Geoinformation, Shijiazhuang 050000, China}

\begin{abstract}
Employing two state-of-the-art methods, multiconfiguration Dirac--Hartree--Fock and second-order many-body perturbation theory, highly accurate calculations are performed for the lowest 272 fine-structure levels arising from the $2s^{2} 2p^{3}$, $2s 2p^{4}$, $2p^{5}$, $2s^{2} 2p^{2} 3l$~($l=s,p,d$), $2s 2p^{3}3l$ ($l=s,p,d$), and $2p^{4} 3l$ ($l=s,p,d$) configurations in nitrogen-like Ge XXVI. 
Complete and consistent atomic data, including excitation energies, lifetimes,  wavelengths, hyperfine structures, Land\'e $g_{J}$-factors, and  E1, E2, M1, M2 line strengths,  oscillator strengths, and transition rates  among these 272 levels are provided. 
Comparisons are made between the present two data sets, as well as with other available experimental and theoretical values. 
The present data are accurate enough for identification and deblending of emission lines involving the $n=3$ levels, and are also useful for modeling and diagnosing fusion plasmas.
\end{abstract}


\end{frontmatter}

\twocolumn
\section{Introduction}
Spectra of N-like ions with $Z =30-36$ have received a great attention both experimentally and theoretically, because of their wide applications in fusion plamas~\citep{Wouters.1988.V5.p1520,Feldman.1989.V6.p1652,Denne.1989.V40.p1488,Kink.2001.V63.p46409,Fournier.2003.V36.p3787,Saloman.2007.V36.p215,Shirai.2007.V36.p509,Podpaly.2014.V47.p95702}. 
Using high-energy lasers or tokamak discharges, spectra of N-like ions, including Zn XXIV, Se XXVIII, and Kr XXX, have been measured in plasmas~\citep{Hinnov.1982.V25.p2293,Hinnov.1986.V3.p1288,Feldman.1986.V3.p1605,Hinnov.1987.V4.p1293,Wouters.1988.V5.p1520,Feldman.1989.V6.p1652,Denne.1989.V40.p1488,Fournier.2003.V36.p3787}. 
In regard to N-like Ge XXVI, two M1 transitions $(1s^2) 2s^2 2p^3~^4\!S_{3/2}~-~^2\!D_{3/2,5/2}$ were identified by~\citet{Denne.1987.V35.p18}. 
~\citet{Behring.1985.V2.p886} observed nine E1 transition lines of the arrays $2s^2 2p^3-2s 2p^4$ and $2s 2p^4-2p^5$. Their results were extended by ~\citet{Feldman.1989.V6.p1652} to 22 lines among the $n = 2$ levels by axially observing a laser produced plasma.

Experiments can, due to limited resources, never provide complete data sets for these N-like ions. Instead, the bulk of the data must be calculated. Theoretical studies  have been performed using different methods~\citep{Zhang.1999.V72.p153,Gu.2005.V89.p267,Han.2014.V89.p42514,Fontes.2014.V100.p1292,Rynkun.2014.V100.p315}, in which excitation energies  and transition rates for the $n = 2$ levels were provided. 
It is clear that atomic data involving the $n > 2$ levels are also important because of their wide applications in plasma physics~\citep{Rice.2000.V33.p5435,Kink.2001.V63.p46409}. 
In view of this, we have provided energy and transition data involving the $n > 2$ levels for ions from Ar XII to Zn XXIV~\citep{Radziute.2015.V582.p61,Wang.2016.V223.p3} and Kr XXX~\citep{Wang.2017.V187.p375}. 
The accuracy of our calculations is high enough to facilitate identifications of spectral lines, and the data are also useful for modeling and diagnosing fusion plasmas. 

This work presents our effort for N-like Ge XXVI to provide the database of energy  and transition data involving high-lying levels. 
Based on the multiconﬁguration Dirac-Hartree-Fock (MCDHF)  and relativistic conﬁguration interaction (RCI) 
methods~\citep{Grant.2007.V.p,FroeseFischer.2016.V49.p182004} implemented in the GRASP2K code~\cite{Jonsson.2007.V177.p597,Jonsson.2013.V184.p2197}, energy levels, wavelengths $\lambda$, line strengths $S$, oscillator strengths $gf$, transition rates $A$, lifetimes $\tau$, hyperfine interaction constants $A_J$ and $B_J$, and Land\'e $g_{J}$-factors are provided here for the 272 levels of the $2s^{2} 2p^{3}$, $2s 2p^{4}$, $2p^{5}$, $2s^{2} 2p^{2} 3l$~($l=s,p,d$), $2s 2p^{3}3l$ ($l=s,p,d$), and $2p^{4} 3l$ ($l=s,p,d$)  configurations. 
To assess the accuracy of the MCDHF/RCI data,  independent calculations are also performed using the many-body perturbation theory (MBPT) method~\cite{Lindgren.1974.V7.p2441,Safronova.1996.V53.p4036,Vilkas.1999.V60.p2808,Gu.2005.V156.p105,Gu.2007.V169.p154} implemented in  the FAC code~\cite{Gu.2008.V86.p675}. 
Comparisons are made with other available experimental and theoretical results, and the accuracy of the present data is assessed. Our calculated energies are accurate enough to directly aid and confirm experimental identifications.
The present work significantly increases the amount of accurate data for the $n = 3$ levels. 

\section{Calculations} 
\subsection{MCDHF} 
The MCDHF method has been described by~\citet{Grant.2007.V.p} and~\citet{FroeseFischer.2016.V49.p182004}.  Based on the active space (AS) approach~\cite{Olsen.1988.V89.p2185,Sturesson.2007.V177.p539} for the generation of the configuration state function (CSF) expansions, separate calculations are done for the even and odd parity states. For the even parity states, the CSF expansions are obtained by allowing single and double (SD) excitations from the multi-reference (MR) configurations $2s 2p^{4}$, $2s^2 2p^2 3s$, $2s^2 2p^2 3d$, $2s 2p^3 3p$, $2p^4 3s$, $2p^4 3d$, $2s^2 2p^2 4s$, and $2s^2 2p^2 4d$  to an AS of orbitals. For the odd parity states, the CSF expansions are obtained by allowing SD excitations from the MR configurations $2s^{2} 2p^{3}$, $2p^{5}$, $2s^2 2p^2 3p$, $2s 2p^3 3s$, $2s 2p^3 3d$,  $2p^4 3p$,  $2s^2 2p^2 4p$, and $2s^2 2p^2 4f$ to an AS of orbitals. In the first step of the calculations, the AS is

$\rm{AS_1 = \{1s, 2s, 2p, 3s, 3p, 3d, 4s, 4p, 4d, 4f\}}$.
\\

Then, the AS is  increased in the following way:

$\rm{AS_2 = AS_1+\{5s, 5p, 5d, 5f, 5g\}}$,
\\

$\rm{AS_3 = AS_2+\{6s, 6p, 6d, 6f, 6g, 6h\}}$,
\\

$\rm{AS_4 = AS_3+\{7s, 7p, 7d, 7f, 7g, 7h\}}$,
\\

$\rm{AS_5 = AS_4+\{8s, 8p, 8d, 8f, 8g, 8h\}}$.
\\

By enlarging the AS layer by layer, the convergence of the computed properties can be monitored. At each stage only the outer orbitals are optimized, while the inner ones are fixed. To reduce the number of CSFs, the $\rm 1s^{2}$ core is closed during the relativistic self-consistent field (RSCF) calculations, but is opened during the RCI calculations, where 
the Breit interaction and leading quantum electrodynamic (QED) effects , i.e., vacuum polarization and self-energy, are included in the Hamiltonian. In the RCI calculations the mixing
coefficients $c_r$ are calculated without changing the radial
functions. The final model using the AS$_5$ active set contains about
1~380~000/7~151~000 even and 1~994~000/10~204~000  odd parity CSFs with the $1s^{2}$ core closed/opened.  Once the atomic state functions (ASFs) have been obtained, atomic parameters, such as line strengths, transition rates, hyperfine interaction constants, and Land\'e $g_{J}$-factors can be calculated. A more detailed description of these parameters can be 
found in our recent work~\citep{Wang.2017.V187.p375} as well as in the original write-ups of the computer codes~\citep{Jonsson.1996.V96.p301,Andersson.2008.V178.p156}. 

\subsection{MBPT}
The MBPT method is explained in~\cite{Lindgren.1974.V7.p2441,Safronova.1996.V53.p4036,Vilkas.1999.V60.p2808,Gu.2005.V156.p105,Gu.2007.V169.p154}.
The method has been implemented in the FAC package~\citep{Gu.2008.V86.p675}, and successfully used to calculate atomic data of high accuracy~\citep{Wang.2014.V215.p26,Wang.2015.V218.p16,Wang.2016.V223.p3,Wang.2016.V226.p14,Wang.2017.V229.p37,Wang.2017.V194.p108,Si.2016.V227.p16,Si.2017.V189.p249,Chen.2017.V113.p258,Chen.2017.V114.p61}. 
The key feature of the MBPT method is the partitioning of the Hilbert space of the system into two subspaces, the model space $M$ and the orthogonal space $O$. The configuration interaction effects in the $M$ space are exactly considered, while the interaction between the space $M$ and $O$ is
taken into account with the second-order perturbation method. 
For the MBPT calculation, the model space $M$ contains the even and odd multi-reference configurations of the MCDHF method,
while the space $O$ contains all the possible configurations
that are generated by SD virtual excitations of the $O$ space.
For single/double excitations, the maximum $n$ value is 125/65, with the maximum $l$ value of 25. Just as for the MCDHF calculations, QED effects are also included.

\section{Results and Discussions}
In the relativistic calculations, the ASFs are obtained as expansions over $jj$-coupled CSFs. To provide the $LSJ$ labeling system used by the experimentalists, as well as used in
other sources, the ASFs are transformed  from a $jj$-coupled CSF basis into an
$LSJ$-coupled CSF basis using the method provided by~\citeauthor{Gaigalas.2004.V157.p239}~\citep{Gaigalas.2004.V157.p239,Gaigalas.2017.V5.p6}. The computed excitation energies for all the 272 levels of the $2s^{2} 2p^{3}$, $2s 2p^{4}$, $2p^{5}$, $2s^{2} 2p^{2} 3l$~($l=s,p,d$), $2s 2p^{3}3l$ ($l=s,p,d$), and $2p^{4} 3l$ ($l=s,p,d$)  configurations from our MCDHF/RCI and MBPT calculations are listed in Table~\ref{tableE}, along with the radiative lifetimes estimated from E1, E2, M1, and M2 transition rates, and the $LSJ$-coupled and $jj$-coupled labels obtained from our calculations. 
Table~\ref{tab:3} lists wavelengths, and E1, E2, M1, and M2 line strengths $S$, oscillator strengths $gf$, and radiative rates $A$ among the 272 energy levels, obtained from both the MCDHF/RCI and MBPT methods. All the E1 and E2 values are computed in the Babushkin gauge (equivalent to the non-relativistic length form), which is considered to be more accurate than the Coulomb gauge (equivalent to the non-relativistic velocity form).

\subsection{Excitation energies}
In Table~\ref{tab:1}, we present the MCDHF excitation energies of the 272 levels  as a function of the increasing active set (AS). 
When the AS is increased from AS$_{k-1}$ to AS$_{k}$, see section 2.1 for the definition of the AS, the energy differences $\Delta E_{k,k-1} \equiv (E_{\rm AS_{k}} 
-E_{\rm AS_{k-1}})$ for each of the  272 levels can be compared.  The average absolute differences between the AS$_{k-1}$ and AS$_{k}$ excitation energies along with the standard deviation are found to be
$-267 \pm 4 182  $ cm$^{-1}$, $-275 \pm
634 $ cm$^{-1}$, $13
 \pm 152 $ cm$^{-1}$, and $25  \pm  61$ cm$^{-1}$ for, respectively, $k=2,3,4,5$. The MCDHF calculations are thus well converged with respect to an increasing size of the AS.   Based on the the AS$_5$ expansion, including the correlation effects from the $1s^2$ electrons, the RCI excitation energies (hereafter referred to as RCI1) are presented in the Table~\ref{tab:1}. The correlation effects from the $1s^2$ electrons, included in the RCI1 calculations, change the excitation energies by amounts ranging from -3 500 cm$^{-1}$ to 1100  cm$^{-1}$ for the AS$_5$ expansion. 
 
\begin{figure}
	\includegraphics[width=20cm]{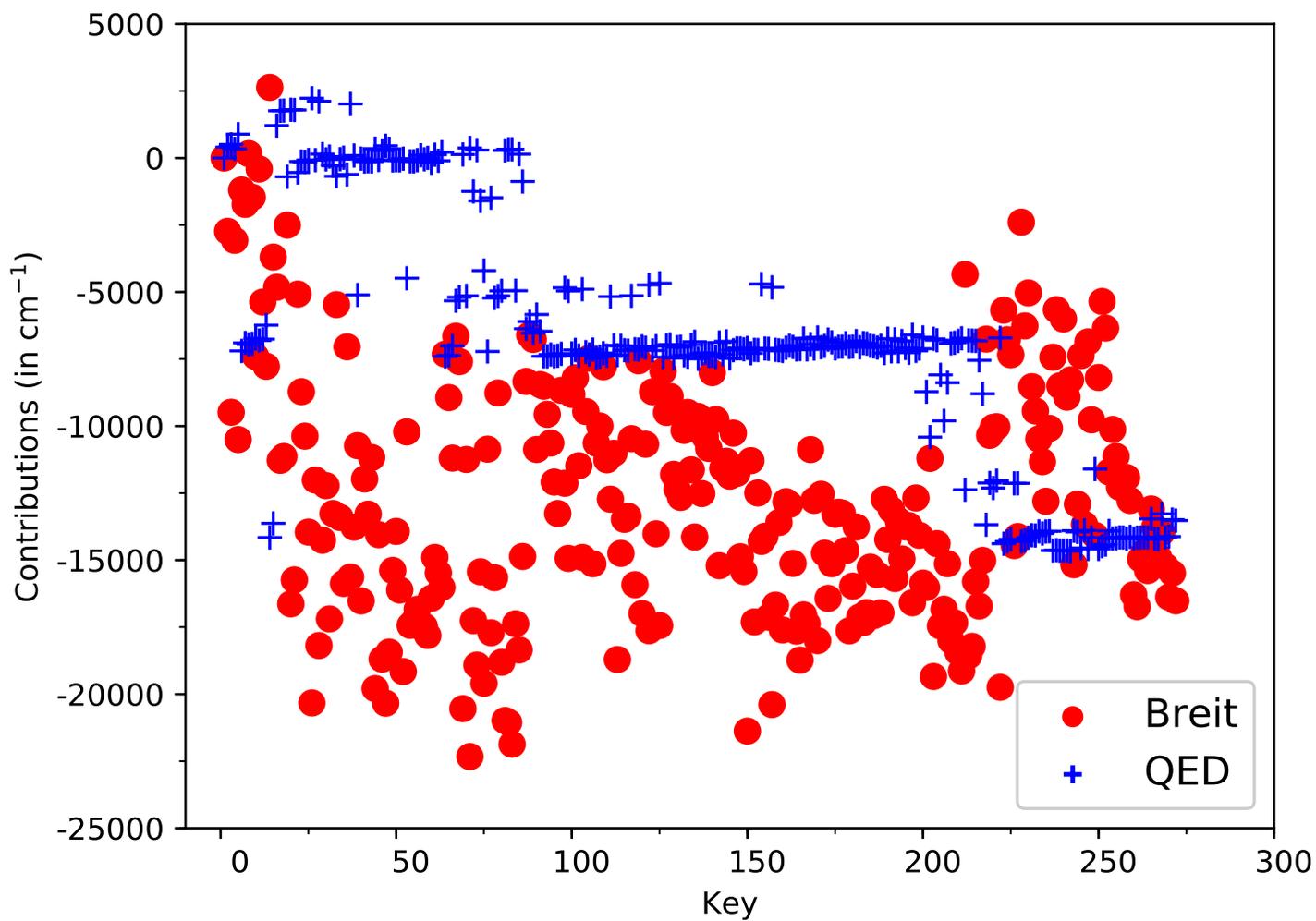}% Here is how to import EPS art
	\caption{\label{fig:0}Contributions (in cm$^{-1}$) of the Breit and QED effects on the MCDHF/RCI2 excitation energies for the 272 levels of Ge XXVI.}
\end{figure}

\begin{figure}
	\includegraphics[width=20cm]{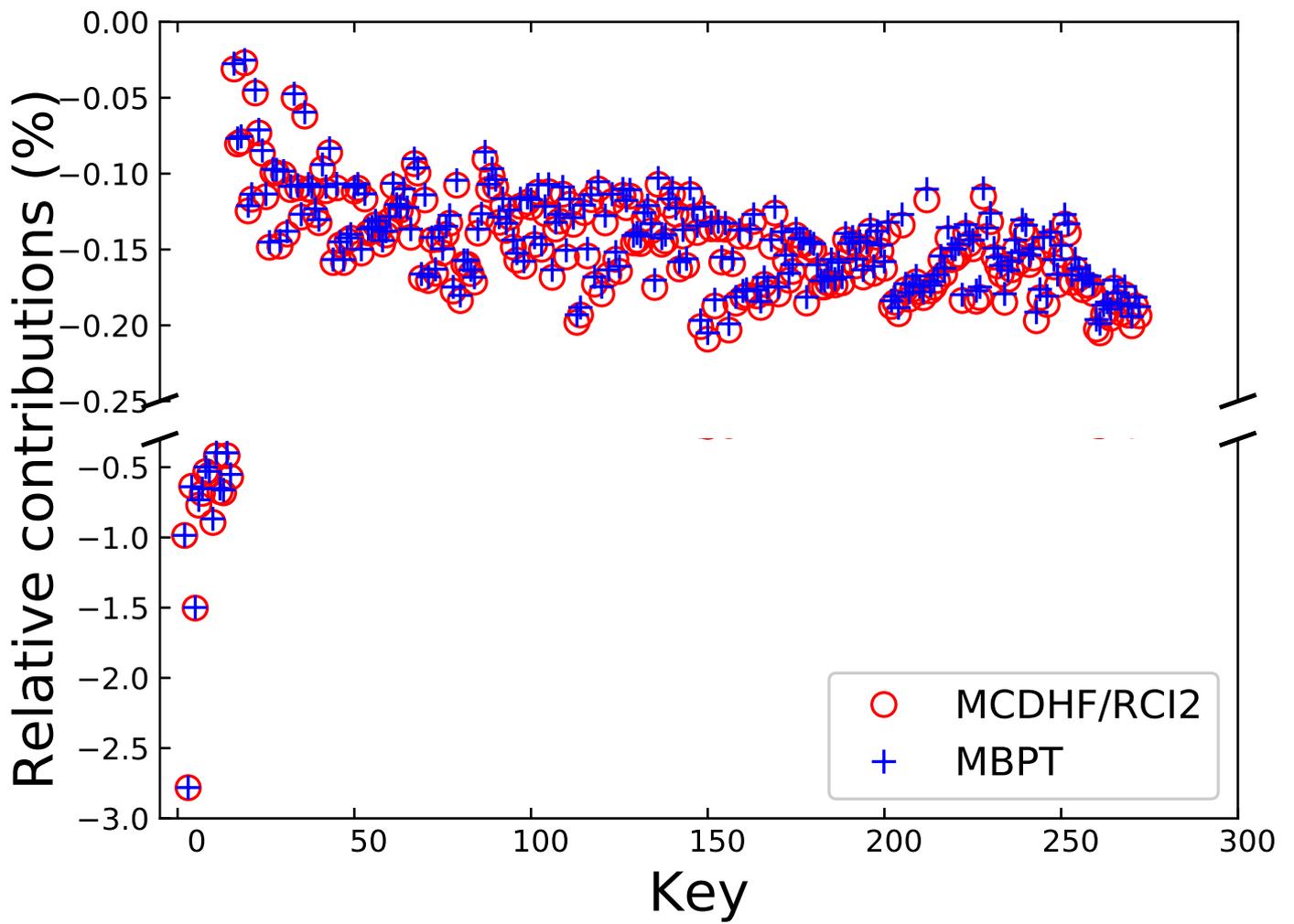}% Here is how to import EPS art
	\caption{\label{fig:1}Relative total contributions (in \%) of the Breit and QED effects on the MCDHF/RCI2 and MBPT excitation energies for the 272 levels of Ge XXVI.}
\end{figure}

Furthermore, the RCI excitation energies (hereafter referred to as RCI2), including both the $1s^2$ electron correlation effects and the Breit and QED effects, are also listed in Table~\ref{tab:1}. By comparing the RCI1 and RCI2 results, it is shown that the Breit and QED effects have contributions ranging from -2 300  cm$^{-1}$ to -31 000 cm$^{-1}$ to excitation energies, which is indispensable for accurate prediction of energy levels. The individual Breit and QED effects are shown in Figure~\ref{fig:0}. It is seen that the  Breit corrections are significant, generally lowering the excited levels. For the lowest 15 levels of the $2s^{2}2p^{3}$, $2s2p^{4}$, and $2p^{5}$ configurations, the Breit results are lower than the corresponding Coulomb excitation energies by about 400 cm$^{-1}$ -11000 cm$^{-1}$  with two exceptions for the $2s\,2p^4 ~^4P_{1/2}$ and $2p^5 ~^2P_{3/2}$ levels, where the former results are higher than the latter values by about 165 cm$^{-1}$ and 2635 cm$^{-1}$, respectively. For the remaining levels belonging to the $2s^{2}2p^{2} nl$ and $2s2p^{3} nl$ ($n= 3$, $l=s,p,d$) configurations, the Breit effects reduce the corresponding Coulomb excitation energies by about -2400 cm$^{-1}$ - 23000 cm$^{-1}$. On the other hand, the QED effects generally  reduce  the excitation energies by up to about -15000 cm$^{-1}$. Moreover, we can see that the QED corrections are naturally grouped according to the number of $s$-orbital electron of the configurations, i.e., the $2s^{2}2p^{3}$, $2s2p^{4}$, $2p^{5}$, $2s^{2}2p^{2}3l$, $2s2p^{3}3l$, and $2p^{4}3l$ groups. The QED effects on the energies of the configurations $2p^{4}3l$  (without $2s$ electron) are generally larger than the configurations $2s2p^{3}3l$  (with one $2s$ electron) by about 6000 cm$^{-1}$ - 8000 cm$^{-1}$. Similarly, their effects on the configurations $2s2p^{3}3l$ (with one $2s$ electron) are generally larger than the configurations $2s^{2}2p^{2}3l$ (with two $2s$ electron) by about 5000 cm$^{-1}$ - 7000 cm$^{-1}$. To more clearly assess the Breit and QED effects, we also use the MBPT method to provide the results, i.e., the MBPT1 values (excluding the Breit and QED effects), and the MBPT2 values (including the Breit and QED effects), which are also included in Table~\ref{tab:1}. As shown in Figure~\ref{fig:1}, their relative contributions (in \%) to the MCDHF/RCI2 and MBPT2 excitation energies of all the 272 levels show good agreement. By including the contributions from the Breit and QED effects, excitation energies of the $n=2$ and $n=3$ levels are reduced by about 0.4~\%-2.8~\% and 0.03~\%-0.21~\%, respectively.
 
 As shown in Table~\ref{tab:1}, the MCDHF/RCI2 and MBPT2 excitation energies are in very good agreement for both the $n=2$ and $n=3$ levels. For the $n=2$ levels, the absolute difference of the two data sets is within 800 cm$^{-1}$. Experimental determinations in the Atomic Spectra Database (ASD) of the National Institute of Standards and Technology (NIST)~\cite{Kramida.2015.V.p}  are available for the $n=2$ levels. The agreement of the NIST values and the present MCDHF/RCI2 (or MBPT2) excitation energies is very good, and the absolute difference is within 600  cm$^{-1}$. For the remaining levels belonging to the $n=3$ configurations, the average absolute difference with the standard deviation of the present MBPT2 and MCDHF/RCI2 excitation energies is $-50 \pm 462$ cm$^{-1}$, corresponding to the average relative difference with the standard deviation of $-0.0005\% \pm 0.004\%$. 

\subsection{Transition rates}\label{sec_tr}
\begin{figure}
%    \centering
    \includegraphics[width=20cm]{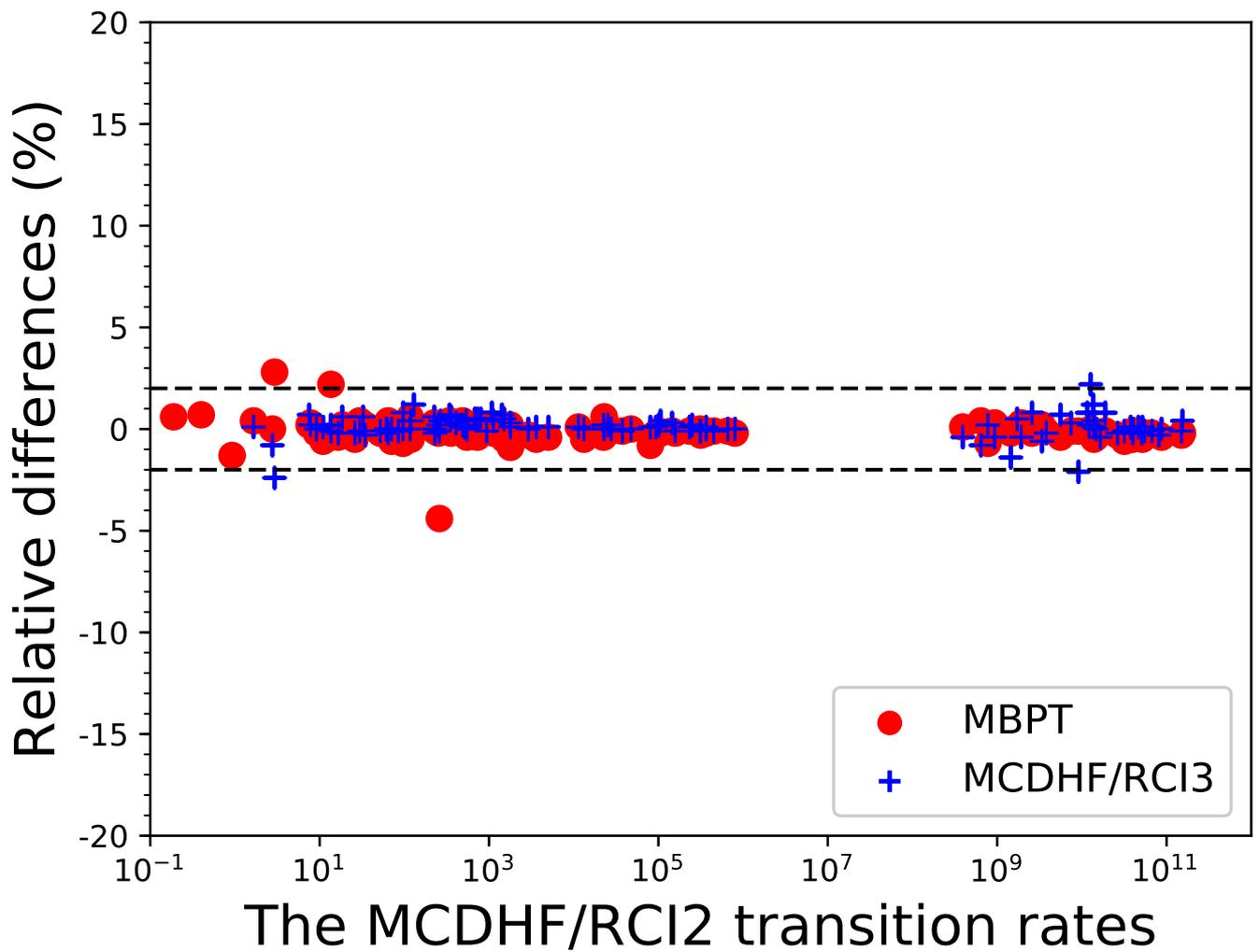}% Here is how to import EPS art
    \caption{\label{fig:2} Differences (in \%) of the MBPT transition rates and MCDHF/RCI3 transition rates of \citet{Rynkun.2014.V100.p315} relative to the present MCDHF/RCI2 values among the 15 levels. Dashed lines indicate the differences of $\pm~2~\%$.}
\end{figure}

Among the calculations~\cite{Fontes.2014.V100.p1292,Rynkun.2014.V100.p315,Gu.2005.V89.p267,Zhang.1999.V72.p153}  of the $n=2$ levels for N-like ions, transition data (hereafter referred to as MCDHF/RCI3) reported by \citet{Rynkun.2014.V100.p315} are the most accurate so far. 
In Figure~\ref{fig:2}, we compare the present two sets of transition rates among the 15 levels belonging to the $n=2$ configurations with the MCDHF/RCI3 values.  
The present two data sets and the MCDHF/RCI3 values are in good agreement, which is within 2~\% for most transitions, with the largest difference of 5~\%. 

According to the uncertainty estimation method suggested by~\citet{Kramida.2014.V2.p22,Kramida.2014.V212.p11}  the averaged uncertainties for the line strengths $S$ of E1 transitions from the present MCDHF/RCI and MBPT calculations in various ranges of $S$ are assessed to be 1.5~\% for $S \geq 10^{-1}$; 4~\% for $10^{-1} > S \geq 10^{-2}$; 6~\% for $10^{-2} > S \geq 10^{-3}$; 9~\% for $10^{-3} > S \geq 10^{-4}$; and 19~\% for $10^{-4} > S \geq 10^{-5}$. Accounting also for the contributions from the uncertainty of the wavelengths, about 2.7~\% of the E1 transitions included in Table~\ref{tab:3} have uncertainties of the transition rate $A$ of $\leq$ 3~\% (the categories A$^{+}$ or A  in the terminology of the NIST ASD~\cite{Kramida.2015.V.p}),  43.1~\% have uncertainties of  $\leq$ 7~\% (the category B$^{+}$), 36.5~\% have uncertainties of  $\leq$ 10~\% (the category B), 2.1~\% have uncertainties of  $\leq$ 18~\% (the category C$^{+}$), 9.1~\% have uncertainties of  $\leq$ 25~\% (the category C), while only 6.4~\% have uncertainties of  $>$ 40~\% (categories D$^{+}$, D, and E). The uncertainty estimates of the $A$ values are listed in the last column of Table~\ref{tab:3}. The largest differences between the two sets of results generally occur for transitions with large cancellation effects~\cite{Si.2016.V227.p16} or weak transitions.  For example, as shown in Table~\ref{tableE}, the levels with index 72 and 74 ($2s^{2}\,2p^{2}(^{1}_{2}D)~^{1}D\,3d~^{2}D_{3/2}$) and $2s~^{2}S\,2p^{3}(^{4}_{3}S)~^{5}S\,3p~^{4}P_{3/2}$) are strongly mixed. The transitions involving these two levels have  large cancellation effects. Even a slight difference in the calculations will lead to a relatively large difference in the computed $S$ and $A$ values, which has been pointed out in our recent work~\cite{Si.2016.V227.p16}.  Most of weak transitions are two-electrons-one-photon transitions. These transitions are strictly forbidden in the single configuration approximation and are induced through configuration interaction effects. 
Even with today's methods, which allow massive CSF expansions, such transitions are very difficult to compute accurately. 

Again, using the method suggested in~\cite{Kramida.2014.V2.p22,Kramida.2014.V212.p11}, the uncertainties of the $A$ values for the M1, E2, and M2 transitions  are estimated. The estimated uncertainties for all M1, E2, and M2 transitions are listed in Table~\ref{tab:3}.

\subsection{Lifetimes, Hyperfine interaction constants, and Land\'e $g_{J}$-factors}
Lifetimes for the lowest 4 excited levels of the $2s^2 2p^3$ configuration are determined by the M1 transitions to the ground state. The lifetimes of the other levels are mostly dominated by E1 transitions, but the E2 transition rates from some $2s^2 2p^2 3p$ levels to the $2s^2 2p^3$ levels, and from some $2p^4 3d$ to the $2s 2p^4$ levels are not negligible, since their contributions may reach up to 20~\%. Our two data sets agree well within 6~\% except for two levels, i.e.,  levels 72 ($2s^{2}\,2p^{2}(^{1}_{2}D)~^{1}D\,3d~^{2}D_{3/2}$) and 74  ($2s~^{2}S\,2p^{3}(^{4}_{3}S)~^{5}S\,3p~^{4}P_{3/2}$), for which the differences are 7~\% and 20~\%, respectively. The large differences  are due to 
the strong mixing of the states as discussed in Section~\ref{sec_tr}.

The total energies, $A_{J}$, $B_{J}$ hyperfine
interaction constants and Land\'e $g_{J}$-factors for the
272 levels of Ge XXVI calculated using the MCDHF/RCI method are also given
in Table~\ref{tab:5}. In the present calculations, the nuclear parameters $I$, $\mu$$_{I}$, and $Q$ are all set to 1. To obtain the $A_J$ and $B_J$ values for a specific isotope, the given values can be scaled with the tabulated values. The only available results for  the $A_{J}$, $B_{J}$ constants and the
Land\'e $g_{J}$-factors are the data for the $n=2$ provided by~\citet{Verdebout.2014.V100.p1111}. The present results for $A_{J}$, $B_{J}$  show good agreement, which is within 2~\%,  with Ref.~\cite{Verdebout.2014.V100.p1111}. The Land\'e $g_{J}$-factors, which are known to be insensitive to electron correlation effects, are essentially identical to  the ones calculated by \citet{Verdebout.2014.V100.p1111}.

\section{Conclusions}
Using the MCDHF/RCI and MBPT methods, energy levels, lifetimes, wavelengths, hyperfine interaction constants,
Land\'e $g_{J}$-factors, E1, M1, E2, and M2 transition rates, line strengths, and
oscillator strengths for the lowest 272 levels belonging to the $2s^{2} 2p^{3}$, $2s 2p^{4}$, $2p^{5}$, $2s^{2} 2p^{2} 3l$~($l=s,p,d$), $2s 2p^{3}3l$ ($l=s,p,d$), and $2p^{4} 3l$ ($l=s,p,d$)  configurations of N-like Ge XXVI have been determined. 
The accuracy of energy levels and transition probabilities is estimated by comparing the MCDHF/RCI and MBPT results with available experimental and theoretical data. Excitation energies are accurate to within 800 cm$^{-1}$ for the $n = 2$ levels. For the $n=3$ levels, the average absolute difference with the standard deviation of our two data sets is only $-50 \pm 462$ cm$^{-1}$. Lifetimes are assessed to be accurate to better than 6~\% for most levels. We believe the present data could serve as benchmarks in future line identiﬁcations, and could make important contributions to modeling and diagnosing fusion plasmas.

\section*{Acknowledgments}
We acknowledge the support of the National Natural Science Foundation of China (Grant No.~11703004, No.~11504421, No.~11674066, and No.~11474034), the Nature Science Foundation of Hebei Province, China (A2017201165), and the Project funded by China Postdoctoral Science Foundation (Grant No. 2016M593019). This work is also supported by the Swedish Research Council under contract 2015-04842, Chinese Association of Atomic and Molecular Data, and Chinese National Fusion Project for ITER No. 2015GB117000. One author (K. W.) expresses his gratitude to the support from
the visiting researcher program at the Fudan University.

\onecolumn
\section*{References}

%\bibliography{ref.bib}
\bibliography{ref.bib}

\begin{thebibliography}{52}
\expandafter\ifx\csname natexlab\endcsname\relax\def\natexlab#1{#1}\fi
\providecommand{\url}[1]{\texttt{#1}}
\providecommand{\href}[2]{#2}
\providecommand{\path}[1]{#1}
\providecommand{\DOIprefix}{doi:}
\providecommand{\ArXivprefix}{arXiv:}
\providecommand{\URLprefix}{URL: }
\providecommand{\Pubmedprefix}{pmid:}
\providecommand{\doi}[1]{\href{http://dx.doi.org/#1}{\path{#1}}}
\providecommand{\Pubmed}[1]{\href{pmid:#1}{\path{#1}}}
\providecommand{\bibinfo}[2]{#2}
\ifx\xfnm\relax \def\xfnm[#1]{\unskip,\space#1}\fi
%Type = Article
\bibitem[{{Wouters} et~al.(1988){Wouters}, {Schwob}, {Suckewer}, {Seely},
  {Feldman}, and {Dave}}]{Wouters.1988.V5.p1520}
\bibinfo{author}{A.~W. {Wouters}}, \bibinfo{author}{J.~L. {Schwob}},
  \bibinfo{author}{S.~{Suckewer}}, \bibinfo{author}{J.~F. {Seely}},
  \bibinfo{author}{U.~{Feldman}}, \bibinfo{author}{J.~H. {Dave}},
  \bibinfo{journal}{J. Opt. Soc. Am. B} \bibinfo{volume}{5}
  (\bibinfo{year}{1988}) \bibinfo{pages}{1520--1527}.
%Type = Article
\bibitem[{{Feldman} et~al.(1989){Feldman}, {Ekberg}, {Brown}, and
  {Seely}}]{Feldman.1989.V6.p1652}
\bibinfo{author}{U.~{Feldman}}, \bibinfo{author}{J.~O. {Ekberg}},
  \bibinfo{author}{C.~M. {Brown}}, \bibinfo{author}{J.~F. {Seely}},
  \bibinfo{journal}{J. Opt. Soc. Am. B} \bibinfo{volume}{6}
  (\bibinfo{year}{1989}) \bibinfo{pages}{1652--1655}.
%Type = Article
\bibitem[{Denne et~al.(1989)Denne, Hinnov, Ramette, and
  Saoutic}]{Denne.1989.V40.p1488}
\bibinfo{author}{B.~Denne}, \bibinfo{author}{E.~Hinnov},
  \bibinfo{author}{J.~Ramette}, \bibinfo{author}{B.~Saoutic},
  \bibinfo{journal}{Phys. Rev. A} \bibinfo{volume}{40} (\bibinfo{year}{1989})
  \bibinfo{pages}{1488--1496}.
%Type = Article
\bibitem[{{Kink} et~al.(2001){Kink}, {Laming}, {Tak{\'a}cs}, {Porto},
  {Gillaspy}, {Silver}, {Schnopper}, {Bandler}, {Barbera}, {Brickhouse},
  {Murray}, {Madden}, {Landis}, {Beeman}, and {Haller}}]{Kink.2001.V63.p46409}
\bibinfo{author}{I.~{Kink}}, \bibinfo{author}{J.~M. {Laming}},
  \bibinfo{author}{E.~{Tak{\'a}cs}}, \bibinfo{author}{J.~V. {Porto}},
  \bibinfo{author}{J.~D. {Gillaspy}}, \bibinfo{author}{E.~{Silver}},
  \bibinfo{author}{H.~{Schnopper}}, \bibinfo{author}{S.~R. {Bandler}},
  \bibinfo{author}{M.~{Barbera}}, \bibinfo{author}{N.~{Brickhouse}},
  \bibinfo{author}{S.~{Murray}}, \bibinfo{author}{N.~{Madden}},
  \bibinfo{author}{D.~{Landis}}, \bibinfo{author}{J.~{Beeman}},
  \bibinfo{author}{E.~E. {Haller}}, \bibinfo{journal}{Phys. Rev. E}
  \bibinfo{volume}{63} (\bibinfo{year}{2001}) \bibinfo{pages}{046409}.
%Type = Article
\bibitem[{{Fournier} et~al.(2003){Fournier}, {Faenov}, {Pikuz}, {Magunov},
  {Skobelev}, {Belyaev}, {Vinogradov}, {Kyrilov}, {Matafonov}, {Flora},
  {Bollanti}, {Di Lazzaro}, {Murra}, {Reale}, {Reale}, {Tomassetti}, {Ritucci},
  {Francucci}, {Martellucci}, and {Petrocelli}}]{Fournier.2003.V36.p3787}
\bibinfo{author}{K.~B. {Fournier}}, \bibinfo{author}{A.~Y. {Faenov}},
  \bibinfo{author}{T.~A. {Pikuz}}, \bibinfo{author}{A.~I. {Magunov}},
  \bibinfo{author}{I.~Y. {Skobelev}}, \bibinfo{author}{V.~S. {Belyaev}},
  \bibinfo{author}{V.~I. {Vinogradov}}, \bibinfo{author}{A.~S. {Kyrilov}},
  \bibinfo{author}{A.~P. {Matafonov}}, \bibinfo{author}{F.~{Flora}},
  \bibinfo{author}{S.~{Bollanti}}, \bibinfo{author}{P.~{Di Lazzaro}},
  \bibinfo{author}{D.~{Murra}}, \bibinfo{author}{A.~{Reale}},
  \bibinfo{author}{L.~{Reale}}, \bibinfo{author}{G.~{Tomassetti}},
  \bibinfo{author}{A.~{Ritucci}}, \bibinfo{author}{M.~{Francucci}},
  \bibinfo{author}{S.~{Martellucci}}, \bibinfo{author}{G.~{Petrocelli}},
  \bibinfo{journal}{J. Phys. B: At. Mol. Opt. Phys.} \bibinfo{volume}{36}
  (\bibinfo{year}{2003}) \bibinfo{pages}{3787--3796}.
%Type = Article
\bibitem[{Saloman(2007)}]{Saloman.2007.V36.p215}
\bibinfo{author}{E.~B. Saloman}, \bibinfo{journal}{J. Phys. Chem. Ref. Data}
  \bibinfo{volume}{36} (\bibinfo{year}{2007}) \bibinfo{pages}{215--386}.
%Type = Article
\bibitem[{Shirai et~al.(2007)Shirai, Reader, Kramida, and
  Sugar}]{Shirai.2007.V36.p509}
\bibinfo{author}{T.~Shirai}, \bibinfo{author}{J.~Reader},
  \bibinfo{author}{A.~E. Kramida}, \bibinfo{author}{J.~Sugar},
  \bibinfo{journal}{J. Phys. Chem. Ref. Data} \bibinfo{volume}{36}
  (\bibinfo{year}{2007}) \bibinfo{pages}{509--615}.
%Type = Article
\bibitem[{{Podpaly} et~al.(2014){Podpaly}, {Gillaspy}, {Reader}, and
  {Ralchenko}}]{Podpaly.2014.V47.p95702}
\bibinfo{author}{Y.~A. {Podpaly}}, \bibinfo{author}{J.~D. {Gillaspy}},
  \bibinfo{author}{J.~{Reader}}, \bibinfo{author}{Y.~{Ralchenko}},
  \bibinfo{journal}{J. Phys. B: At. Mol. Opt. Phys.} \bibinfo{volume}{47}
  (\bibinfo{year}{2014}) \bibinfo{pages}{095702}.
%Type = Article
\bibitem[{Hinnov et~al.(1982)Hinnov, Suckewer, Cohen, and
  Sato}]{Hinnov.1982.V25.p2293}
\bibinfo{author}{E.~Hinnov}, \bibinfo{author}{S.~Suckewer},
  \bibinfo{author}{S.~Cohen}, \bibinfo{author}{K.~Sato},
  \bibinfo{journal}{Phys. Rev. A} \bibinfo{volume}{25} (\bibinfo{year}{1982})
  \bibinfo{pages}{2293--2301}.
%Type = Article
\bibitem[{Hinnov et~al.(1986)Hinnov, Boody, Cohen, Feldman, Hosea, Sato,
  Schwob, Suckewer, and Wouters}]{Hinnov.1986.V3.p1288}
\bibinfo{author}{E.~Hinnov}, \bibinfo{author}{F.~Boody},
  \bibinfo{author}{S.~Cohen}, \bibinfo{author}{U.~Feldman},
  \bibinfo{author}{J.~Hosea}, \bibinfo{author}{K.~Sato}, \bibinfo{author}{J.~L.
  Schwob}, \bibinfo{author}{S.~Suckewer}, \bibinfo{author}{A.~Wouters},
  \bibinfo{journal}{J. Opt. Soc. Am. B} \bibinfo{volume}{3}
  (\bibinfo{year}{1986}) \bibinfo{pages}{1288}.
%Type = Article
\bibitem[{Feldman et~al.(1986)Feldman, Richardson, Behring, Reader, Seely,
  Brown, and Ekberg}]{Feldman.1986.V3.p1605}
\bibinfo{author}{U.~Feldman}, \bibinfo{author}{M.~C. Richardson},
  \bibinfo{author}{W.~E. Behring}, \bibinfo{author}{J.~Reader},
  \bibinfo{author}{J.~F. Seely}, \bibinfo{author}{C.~M. Brown},
  \bibinfo{author}{J.~O. Ekberg}, \bibinfo{journal}{J. Opt. Soc. Am. B}
  \bibinfo{volume}{3} (\bibinfo{year}{1986}) \bibinfo{pages}{1605}.
%Type = Article
\bibitem[{Hinnov et~al.(1987)Hinnov, Ramsey, Stratton, Cohen, and
  Timberlake}]{Hinnov.1987.V4.p1293}
\bibinfo{author}{E.~Hinnov}, \bibinfo{author}{A.~Ramsey},
  \bibinfo{author}{B.~Stratton}, \bibinfo{author}{S.~Cohen},
  \bibinfo{author}{J.~Timberlake}, \bibinfo{journal}{J. Opt. Soc. Am. B}
  \bibinfo{volume}{4} (\bibinfo{year}{1987}) \bibinfo{pages}{1293}.
%Type = Article
\bibitem[{Denne and Hinnov(1987)}]{Denne.1987.V35.p18}
\bibinfo{author}{B.~Denne}, \bibinfo{author}{E.~Hinnov},
  \bibinfo{journal}{Phys. Scr.} \bibinfo{volume}{35} (\bibinfo{year}{1987})
  \bibinfo{pages}{811閳ワ拷18}.
%Type = Article
\bibitem[{Behring et~al.(1985)Behring, Seely, Goldsmith, Cohen, Richardson, and
  Feldman}]{Behring.1985.V2.p886}
\bibinfo{author}{W.~E. Behring}, \bibinfo{author}{J.~F. Seely},
  \bibinfo{author}{S.~Goldsmith}, \bibinfo{author}{L.~Cohen},
  \bibinfo{author}{M.~Richardson}, \bibinfo{author}{U.~Feldman},
  \bibinfo{journal}{J. Opt. Soc. Am. B} \bibinfo{volume}{2}
  (\bibinfo{year}{1985}) \bibinfo{pages}{886--890}.
%Type = Article
\bibitem[{{Zhang} and {Sampson}(1999)}]{Zhang.1999.V72.p153}
\bibinfo{author}{H.~L. {Zhang}}, \bibinfo{author}{D.~H. {Sampson}},
  \bibinfo{journal}{At. Data Nucl. Data Tables} \bibinfo{volume}{72}
  (\bibinfo{year}{1999}) \bibinfo{pages}{153}.
%Type = Article
\bibitem[{Gu(2005)}]{Gu.2005.V89.p267}
\bibinfo{author}{M.~F. Gu}, \bibinfo{journal}{At. Data Nucl. Data Tables}
  \bibinfo{volume}{89} (\bibinfo{year}{2005}) \bibinfo{pages}{267 -- 293}.
%Type = Article
\bibitem[{{Han} et~al.(2014){Han}, {Gao}, {Zeng}, {Jin}, {Yan}, and
  {Li}}]{Han.2014.V89.p42514}
\bibinfo{author}{X.~{Han}}, \bibinfo{author}{X.~{Gao}},
  \bibinfo{author}{D.~{Zeng}}, \bibinfo{author}{R.~{Jin}},
  \bibinfo{author}{J.~{Yan}}, \bibinfo{author}{J.~{Li}},
  \bibinfo{journal}{Phys. Rev. A} \bibinfo{volume}{89} (\bibinfo{year}{2014})
  \bibinfo{pages}{042514}.
%Type = Article
\bibitem[{{Fontes} and {Zhang}(2014)}]{Fontes.2014.V100.p1292}
\bibinfo{author}{C.~J. {Fontes}}, \bibinfo{author}{H.~L. {Zhang}},
  \bibinfo{journal}{At. Data Nucl. Data Tables} \bibinfo{volume}{100}
  (\bibinfo{year}{2014}) \bibinfo{pages}{1292--1321}.
%Type = Article
\bibitem[{{Rynkun} et~al.(2014){Rynkun}, {J{\"o}nsson}, {Gaigalas}, and {Froese
  Fischer}}]{Rynkun.2014.V100.p315}
\bibinfo{author}{P.~{Rynkun}}, \bibinfo{author}{P.~{J{\"o}nsson}},
  \bibinfo{author}{G.~{Gaigalas}}, \bibinfo{author}{C.~{Froese Fischer}},
  \bibinfo{journal}{At. Data Nucl. Data Tables} \bibinfo{volume}{100}
  (\bibinfo{year}{2014}) \bibinfo{pages}{315--402}.
%Type = Article
\bibitem[{{Rice} et~al.(2000){Rice}, {Fournier}, {Goetz}, {Marmar}, and
  {Terry}}]{Rice.2000.V33.p5435}
\bibinfo{author}{J.~E. {Rice}}, \bibinfo{author}{K.~B. {Fournier}},
  \bibinfo{author}{J.~A. {Goetz}}, \bibinfo{author}{E.~S. {Marmar}},
  \bibinfo{author}{J.~L. {Terry}}, \bibinfo{journal}{J. Phys. B: At. Mol. Opt.
  Phys.} \bibinfo{volume}{33} (\bibinfo{year}{2000})
  \bibinfo{pages}{5435--5462}.
%Type = Article
\bibitem[{Rad{\v{z}}i{\={u}}t{\.{e}} et~al.(2015)Rad{\v{z}}i{\={u}}t{\.{e}},
  Ekman, J{\"o}nsson, and Gaigalas}]{Radziute.2015.V582.p61}
\bibinfo{author}{L.~Rad{\v{z}}i{\={u}}t{\.{e}}}, \bibinfo{author}{J.~Ekman},
  \bibinfo{author}{P.~J{\"o}nsson}, \bibinfo{author}{G.~Gaigalas},
  \bibinfo{journal}{Astron. {\&} Astrophy.} \bibinfo{volume}{582}
  (\bibinfo{year}{2015}) \bibinfo{pages}{A61}.
%Type = Article
\bibitem[{Wang et~al.(2016)Wang, Si, Dang, J\"onsson, Guo, Li, Chen, Zhang,
  Long, Liu, Li, Hutton, Chen, and Yan}]{Wang.2016.V223.p3}
\bibinfo{author}{K.~Wang}, \bibinfo{author}{R.~Si}, \bibinfo{author}{W.~Dang},
  \bibinfo{author}{P.~J\"onsson}, \bibinfo{author}{X.~L. Guo},
  \bibinfo{author}{S.~Li}, \bibinfo{author}{Z.~B. Chen},
  \bibinfo{author}{H.~Zhang}, \bibinfo{author}{F.~Y. Long},
  \bibinfo{author}{H.~T. Liu}, \bibinfo{author}{D.~F. Li},
  \bibinfo{author}{R.~Hutton}, \bibinfo{author}{C.~Y. Chen},
  \bibinfo{author}{J.~Yan}, \bibinfo{journal}{Astrophy. J. Supp. Ser.}
  \bibinfo{volume}{223} (\bibinfo{year}{2016}) \bibinfo{pages}{3}.
%Type = Article
\bibitem[{Wang et~al.(2017)Wang, Li, J\"onsson, Fu, Dang, Guo, Chen, Yan, Chen,
  and Si}]{Wang.2017.V187.p375}
\bibinfo{author}{K.~Wang}, \bibinfo{author}{S.~Li},
  \bibinfo{author}{P.~J\"onsson}, \bibinfo{author}{N.~Fu},
  \bibinfo{author}{W.~Dang}, \bibinfo{author}{X.~L. Guo},
  \bibinfo{author}{C.~Y. Chen}, \bibinfo{author}{J.~Yan},
  \bibinfo{author}{Z.~B. Chen}, \bibinfo{author}{R.~Si}, \bibinfo{journal}{J.
  Quant. Spectrosc. Radiat. Transf.} \bibinfo{volume}{187}
  (\bibinfo{year}{2017}) \bibinfo{pages}{375 -- 402}.
%Type = Book
\bibitem[{{Grant}(2007)}]{Grant.2007.V.p}
\bibinfo{author}{I.~P. {Grant}}, \bibinfo{title}{{Relativistic Quantum Theory
  of Atoms and Molecules}}, \bibinfo{year}{2007}.
  \DOIprefix\doi{10.1007/978-0-387-35069-1}.
%Type = Article
\bibitem[{{Froese Fischer} et~al.(2016){Froese Fischer}, {Godefroid}, {Brage},
  {J{\"o}nsson}, and {Gaigalas}}]{FroeseFischer.2016.V49.p182004}
\bibinfo{author}{C.~{Froese Fischer}}, \bibinfo{author}{M.~{Godefroid}},
  \bibinfo{author}{T.~{Brage}}, \bibinfo{author}{P.~{J{\"o}nsson}},
  \bibinfo{author}{G.~{Gaigalas}}, \bibinfo{journal}{J. Phys. B: At. Mol. Opt.
  Phys.} \bibinfo{volume}{49} (\bibinfo{year}{2016}) \bibinfo{pages}{182004}.
%Type = Article
\bibitem[{J\"{o}nsson et~al.(2007)J\"{o}nsson, He, {Froese Fischer}, and
  Grant}]{Jonsson.2007.V177.p597}
\bibinfo{author}{P.~J\"{o}nsson}, \bibinfo{author}{X.~He},
  \bibinfo{author}{C.~{Froese Fischer}}, \bibinfo{author}{I.~Grant},
  \bibinfo{journal}{Comput. Phys. Commun.} \bibinfo{volume}{177}
  (\bibinfo{year}{2007}) \bibinfo{pages}{597 -- 622}.
%Type = Article
\bibitem[{{J{\"o}nsson} et~al.(2013){J{\"o}nsson}, {Gaigalas}, {Biero{\'n}},
  {Froese Fischer}, and {Grant}}]{Jonsson.2013.V184.p2197}
\bibinfo{author}{P.~{J{\"o}nsson}}, \bibinfo{author}{G.~{Gaigalas}},
  \bibinfo{author}{J.~{Biero{\'n}}}, \bibinfo{author}{C.~{Froese Fischer}},
  \bibinfo{author}{I.~P. {Grant}}, \bibinfo{journal}{Comput. Phys. Commun.}
  \bibinfo{volume}{184} (\bibinfo{year}{2013}) \bibinfo{pages}{2197--2203}.
%Type = Article
\bibitem[{Lindgren(1974)}]{Lindgren.1974.V7.p2441}
\bibinfo{author}{I.~Lindgren}, \bibinfo{journal}{J. Phys. B: At. Mol. Opt.
  Phys.} \bibinfo{volume}{7} (\bibinfo{year}{1974}) \bibinfo{pages}{2441}.
%Type = Article
\bibitem[{Safronova et~al.(1996)Safronova, Johnson, and
  Safronova}]{Safronova.1996.V53.p4036}
\bibinfo{author}{M.~S. Safronova}, \bibinfo{author}{W.~R. Johnson},
  \bibinfo{author}{U.~I. Safronova}, \bibinfo{journal}{Phys. Rev. A}
  \bibinfo{volume}{53} (\bibinfo{year}{1996}) \bibinfo{pages}{4036--4053}.
%Type = Article
\bibitem[{Vilkas et~al.(1999)Vilkas, Ishikawa, and Koc}]{Vilkas.1999.V60.p2808}
\bibinfo{author}{M.~J. Vilkas}, \bibinfo{author}{Y.~Ishikawa},
  \bibinfo{author}{K.~Koc}, \bibinfo{journal}{Phys. Rev. A}
  \bibinfo{volume}{60} (\bibinfo{year}{1999}) \bibinfo{pages}{2808--2821}.
%Type = Article
\bibitem[{Gu(2005)}]{Gu.2005.V156.p105}
\bibinfo{author}{M.~F. Gu}, \bibinfo{journal}{Astrophy. J. Supp. Ser.}
  \bibinfo{volume}{156} (\bibinfo{year}{2005}) \bibinfo{pages}{105}.
%Type = Article
\bibitem[{Gu(2007)}]{Gu.2007.V169.p154}
\bibinfo{author}{M.~F. Gu}, \bibinfo{journal}{Astrophy. J. Supp. Ser.}
  \bibinfo{volume}{169} (\bibinfo{year}{2007}) \bibinfo{pages}{154}.
%Type = Article
\bibitem[{{Gu}(2008)}]{Gu.2008.V86.p675}
\bibinfo{author}{M.~F. {Gu}}, \bibinfo{journal}{Can. J. Phys.}
  \bibinfo{volume}{86} (\bibinfo{year}{2008}) \bibinfo{pages}{675--689}.
%Type = Article
\bibitem[{{Olsen} et~al.(1988){Olsen}, {Roos}, {J{\o}rgensen}, and
  {Jensen}}]{Olsen.1988.V89.p2185}
\bibinfo{author}{J.~{Olsen}}, \bibinfo{author}{B.~O. {Roos}},
  \bibinfo{author}{P.~{J{\o}rgensen}}, \bibinfo{author}{H.~J.~A. {Jensen}},
  \bibinfo{journal}{J. Chem. Phys.} \bibinfo{volume}{89} (\bibinfo{year}{1988})
  \bibinfo{pages}{2185--2192}.
%Type = Article
\bibitem[{{Sturesson} et~al.(2007){Sturesson}, {J{\"o}nsson}, and {Froese
  Fischer}}]{Sturesson.2007.V177.p539}
\bibinfo{author}{L.~{Sturesson}}, \bibinfo{author}{P.~{J{\"o}nsson}},
  \bibinfo{author}{C.~{Froese Fischer}}, \bibinfo{journal}{Comput. Phys.
  Commun.} \bibinfo{volume}{177} (\bibinfo{year}{2007})
  \bibinfo{pages}{539--550}.
%Type = Article
\bibitem[{{J{\"o}nsson} et~al.(1996){J{\"o}nsson}, {Parpia}, and {Froese
  Fischer}}]{Jonsson.1996.V96.p301}
\bibinfo{author}{P.~{J{\"o}nsson}}, \bibinfo{author}{F.~A. {Parpia}},
  \bibinfo{author}{C.~{Froese Fischer}}, \bibinfo{journal}{Comput. Phys.
  Commun.} \bibinfo{volume}{96} (\bibinfo{year}{1996})
  \bibinfo{pages}{301--310}.
%Type = Article
\bibitem[{{Andersson} and {J{\"o}nsson}(2008)}]{Andersson.2008.V178.p156}
\bibinfo{author}{M.~{Andersson}}, \bibinfo{author}{P.~{J{\"o}nsson}},
  \bibinfo{journal}{Comput. Phys. Commun.} \bibinfo{volume}{178}
  (\bibinfo{year}{2008}) \bibinfo{pages}{156--170}.
%Type = Article
\bibitem[{Wang et~al.(2014)Wang, Li, Liu, Han, Duan, Li, Li, Guo, Chen, and
  Yan}]{Wang.2014.V215.p26}
\bibinfo{author}{K.~Wang}, \bibinfo{author}{D.~F. Li}, \bibinfo{author}{H.~T.
  Liu}, \bibinfo{author}{X.~Y. Han}, \bibinfo{author}{B.~Duan},
  \bibinfo{author}{C.~Y. Li}, \bibinfo{author}{J.~G. Li},
  \bibinfo{author}{X.~L. Guo}, \bibinfo{author}{C.~Y. Chen},
  \bibinfo{author}{J.~Yan}, \bibinfo{journal}{Astrophy. J. Supp. Ser.}
  \bibinfo{volume}{215} (\bibinfo{year}{2014}) \bibinfo{pages}{26}.
%Type = Article
\bibitem[{Wang et~al.(2015)Wang, Guo, Liu, Li, Long, Han, Duan, Li, Huang,
  Wang, Hutton, Zou, Zeng, Chen, and Yan}]{Wang.2015.V218.p16}
\bibinfo{author}{K.~Wang}, \bibinfo{author}{X.~L. Guo}, \bibinfo{author}{H.~T.
  Liu}, \bibinfo{author}{D.~F. Li}, \bibinfo{author}{F.~Y. Long},
  \bibinfo{author}{X.~Y. Han}, \bibinfo{author}{B.~Duan},
  \bibinfo{author}{J.~G. Li}, \bibinfo{author}{M.~Huang},
  \bibinfo{author}{Y.~S. Wang}, \bibinfo{author}{R.~Hutton},
  \bibinfo{author}{Y.~M. Zou}, \bibinfo{author}{J.~L. Zeng},
  \bibinfo{author}{C.~Y. Chen}, \bibinfo{author}{J.~Yan},
  \bibinfo{journal}{Astrophy. J. Supp. Ser.} \bibinfo{volume}{218}
  (\bibinfo{year}{2015}) \bibinfo{pages}{16}.
%Type = Article
\bibitem[{Wang et~al.(2016)Wang, Chen, Si, J{\"o}nsson, Ekman, Guo, Li, Long,
  Dang, Zhao, Hutton, Chen, Yan, and Yang}]{Wang.2016.V226.p14}
\bibinfo{author}{K.~Wang}, \bibinfo{author}{Z.~B. Chen},
  \bibinfo{author}{R.~Si}, \bibinfo{author}{P.~J{\"o}nsson},
  \bibinfo{author}{J.~Ekman}, \bibinfo{author}{X.~L. Guo},
  \bibinfo{author}{S.~Li}, \bibinfo{author}{F.~Y. Long},
  \bibinfo{author}{W.~Dang}, \bibinfo{author}{X.~H. Zhao},
  \bibinfo{author}{R.~Hutton}, \bibinfo{author}{C.~Y. Chen},
  \bibinfo{author}{J.~Yan}, \bibinfo{author}{X.~Yang},
  \bibinfo{journal}{Astrophy. J. Supp. Ser.} \bibinfo{volume}{226}
  (\bibinfo{year}{2016}) \bibinfo{pages}{14}.
%Type = Article
\bibitem[{Wang et~al.(2017{\natexlab{a}})Wang, J\"onsson, Ekman, Gaigalas,
  Godefroid, Si, Chen, Li, Chen, and Yan}]{Wang.2017.V229.p37}
\bibinfo{author}{K.~Wang}, \bibinfo{author}{P.~J\"onsson},
  \bibinfo{author}{J.~Ekman}, \bibinfo{author}{G.~Gaigalas},
  \bibinfo{author}{M.~R. Godefroid}, \bibinfo{author}{R.~Si},
  \bibinfo{author}{Z.~B. Chen}, \bibinfo{author}{S.~Li}, \bibinfo{author}{C.~Y.
  Chen}, \bibinfo{author}{J.~Yan}, \bibinfo{journal}{Astrophy. J. Supp. Ser.}
  \bibinfo{volume}{229} (\bibinfo{year}{2017}{\natexlab{a}})
  \bibinfo{pages}{37}.
%Type = Article
\bibitem[{Wang et~al.(2017{\natexlab{b}})Wang, J\"onsson, Ekman, Si, Chen, Li,
  Chen, and Yan}]{Wang.2017.V194.p108}
\bibinfo{author}{K.~Wang}, \bibinfo{author}{P.~J\"onsson},
  \bibinfo{author}{J.~Ekman}, \bibinfo{author}{R.~Si},
  \bibinfo{author}{Z.~Chen}, \bibinfo{author}{Y.~Li},
  \bibinfo{author}{C.~Chen}, \bibinfo{author}{J.~Yan}, \bibinfo{journal}{J.
  Quant. Spectrosc. Radiat. Transf.} \bibinfo{volume}{194}
  (\bibinfo{year}{2017}{\natexlab{b}}) \bibinfo{pages}{108--112}.
%Type = Article
\bibitem[{Si et~al.(2016)Si, Li, Guo, Chen, Brage, J{\"o}nsson, Wang, Yan,
  Chen, and Zou}]{Si.2016.V227.p16}
\bibinfo{author}{R.~Si}, \bibinfo{author}{S.~Li}, \bibinfo{author}{X.~L. Guo},
  \bibinfo{author}{Z.~B. Chen}, \bibinfo{author}{T.~Brage},
  \bibinfo{author}{P.~J{\"o}nsson}, \bibinfo{author}{K.~Wang},
  \bibinfo{author}{J.~Yan}, \bibinfo{author}{C.~Y. Chen},
  \bibinfo{author}{Y.~M. Zou}, \bibinfo{journal}{Astrophy. J. Supp. Ser.}
  \bibinfo{volume}{227} (\bibinfo{year}{2016}) \bibinfo{pages}{16}.
%Type = Article
\bibitem[{Si et~al.(2017)Si, Zhang, Liu, Chen, Guo, Li, Yan, Chen, and
  Wang}]{Si.2017.V189.p249}
\bibinfo{author}{R.~Si}, \bibinfo{author}{C.~Zhang}, \bibinfo{author}{Y.~Liu},
  \bibinfo{author}{Z.~Chen}, \bibinfo{author}{X.~Guo}, \bibinfo{author}{S.~Li},
  \bibinfo{author}{J.~Yan}, \bibinfo{author}{C.~Chen},
  \bibinfo{author}{K.~Wang}, \bibinfo{journal}{J. Quant. Spectrosc. Radiat.
  Transf.} \bibinfo{volume}{189} (\bibinfo{year}{2017})
  \bibinfo{pages}{249--257}.
%Type = Article
\bibitem[{Chen et~al.(2017)Chen, Ma, Wang, Wang, Liu, and
  Zeng}]{Chen.2017.V113.p258}
\bibinfo{author}{Z.~B. Chen}, \bibinfo{author}{K.~Ma}, \bibinfo{author}{H.~J.
  Wang}, \bibinfo{author}{K.~Wang}, \bibinfo{author}{X.~B. Liu},
  \bibinfo{author}{J.~L. Zeng}, \bibinfo{journal}{At. Data Nucl. Data Tables}
  \bibinfo{volume}{113} (\bibinfo{year}{2017}) \bibinfo{pages}{258--292}.
%Type = Article
\bibitem[{Chen and Wang(2017)}]{Chen.2017.V114.p61}
\bibinfo{author}{Z.~B. Chen}, \bibinfo{author}{K.~Wang}, \bibinfo{journal}{At.
  Data Nucl. Data Tables} \bibinfo{volume}{114} (\bibinfo{year}{2017})
  \bibinfo{pages}{61--261}.
%Type = Article
\bibitem[{{Gaigalas} et~al.(2004){Gaigalas}, {Zalandauskas}, and
  {Fritzsche}}]{Gaigalas.2004.V157.p239}
\bibinfo{author}{G.~{Gaigalas}}, \bibinfo{author}{T.~{Zalandauskas}},
  \bibinfo{author}{S.~{Fritzsche}}, \bibinfo{journal}{Comput. Phys. Commun.}
  \bibinfo{volume}{157} (\bibinfo{year}{2004}) \bibinfo{pages}{239--253}.
%Type = Article
\bibitem[{Gaigalas et~al.(2017)Gaigalas, {Froese Fischer}, Rynkun, and
  J\"onsson}]{Gaigalas.2017.V5.p6}
\bibinfo{author}{G.~Gaigalas}, \bibinfo{author}{C.~{Froese Fischer}},
  \bibinfo{author}{P.~Rynkun}, \bibinfo{author}{P.~J\"onsson},
  \bibinfo{journal}{Atoms} \bibinfo{volume}{5} (\bibinfo{year}{2017})
  \bibinfo{pages}{6}.
%Type = Misc
\bibitem[{Kramida et~al.(2016)Kramida, {Yu.~Ralchenko}, Reader, and {and NIST
  ASD Team}}]{Kramida.2015.V.p}
\bibinfo{author}{A.~Kramida}, \bibinfo{author}{{Yu.~Ralchenko}},
  \bibinfo{author}{J.~Reader}, \bibinfo{author}{{and NIST ASD Team}},
  \bibinfo{howpublished}{{NIST Atomic Spectra Database (ver. 5.4), [Online].
  Available: {\tt{http://physics.nist.gov/asd}} [2016, May 10]. National
  Institute of Standards and Technology, Gaithersburg, MD.}},
  \bibinfo{year}{2016}.
%Type = Article
\bibitem[{Kramida(2014)}]{Kramida.2014.V2.p22}
\bibinfo{author}{A.~Kramida}, \bibinfo{journal}{Atoms} \bibinfo{volume}{2}
  (\bibinfo{year}{2014}) \bibinfo{pages}{86--122}.
%Type = Article
\bibitem[{{Kramida}(2014)}]{Kramida.2014.V212.p11}
\bibinfo{author}{A.~{Kramida}}, \bibinfo{journal}{Astrophy. J. Supp. Ser.}
  \bibinfo{volume}{212} (\bibinfo{year}{2014}) \bibinfo{pages}{11}.
%Type = Article
\bibitem[{{Verdebout} et~al.(2014){Verdebout}, {Naz{\'e}}, {J{\"o}nsson},
  {Rynkun}, {Godefroid}, and {Gaigalas}}]{Verdebout.2014.V100.p1111}
\bibinfo{author}{S.~{Verdebout}}, \bibinfo{author}{C.~{Naz{\'e}}},
  \bibinfo{author}{P.~{J{\"o}nsson}}, \bibinfo{author}{P.~{Rynkun}},
  \bibinfo{author}{M.~{Godefroid}}, \bibinfo{author}{G.~{Gaigalas}},
  \bibinfo{journal}{At. Data Nucl. Data Tables} \bibinfo{volume}{100}
  (\bibinfo{year}{2014}) \bibinfo{pages}{1111--1155}.

\end{thebibliography}
%|
%\end{comment}
\onecolumn
%\section*{Figures}
%\clearpage

\clearpage

\clearpage

%\section*{Tables}
\newpage
\linespread{1}
\scriptsize
\setlength{\tabcolsep}{5pt}
\begin{longtable}{clrrrrrl}
	\caption{\label{tableE}Energies ($E$ in cm$^{-1}$) relative to the ground state for the lowest  272 levels arising from the $2s^{2} 2p^{3}$, $2s 2p^{4}$, $2p^{5}$, $2s^{2} 2p^{2} 3l$~($l=s,p,d$), $2s 2p^{3}3l$ ($l=s,p,d$), and $2p^{4} 3l$ ($l=s,p,d$) configurations of Ge XXVI. }\\
	\toprule
	Key & $LSJ$-coupled CSF& $E_{\rm NIST}$ & $E_{\rm MCDHF/RCI}$  & $E_{\rm MBPT}$ & $\tau_{\rm MCDHF/RCI}$ & $\tau_{\rm MBPT}$ & $jj$-coupled CSF$^{a,b,c}$\\
	\midrule
	\endfirsthead
	\caption{(\emph{continued})} \\
	\toprule
	Key & $LSJ$-coupled CSF& $E_{\rm NIST}$ & $E_{\rm MCDHF/RCI}$  & $E_{\rm MBPT}$ & $\tau_{\rm MCDHF/RCI}$ & $\tau_{\rm MBPT}$ & $jj$-coupled CSF$^{a,b,c}$\\	\midrule
	\endhead
	\endfoot
	\bottomrule
	\endlastfoot
1   & $2s^{2}\,2p^{3}(^{4}_{3}S)~^{4}S_{3/2}^{\circ}$                &       0& 0         &     0        &               &                  &  2p+1(3)3                                  \\
2   & $2s^{2}\,2p^{3}(^{2}_{3}D)~^{2}D_{3/2}^{\circ}$                &  233740& 233859    &     233805   &      4.30E-06 &        4.30E-06  &  2p-1(1)1 2p+2(4)3                         \\
3   & $2s^{2}\,2p^{3}(^{2}_{3}D)~^{2}D_{5/2}^{\circ}$                &  313520& 313767    &     313636   &      2.44E-05 &        2.44E-05  &  2p-1(1)1 2p+2(4)5                         \\
4   & $2s^{2}\,2p^{3}(^{2}_{1}P)~^{2}P_{1/2}^{\circ}$                &  426510& 426704    &     426572   &      2.62E-06 &        2.63E-06  &  2p-1(1)1 2p+2(0)1                         \\
5   & $2s^{2}\,2p^{3}(^{2}_{1}P)~^{2}P_{3/2}^{\circ}$                &  629730& 630081    &     630053   &      1.07E-06 &        1.07E-06  &  2p+3(3)3                                  \\
6   & $2s~^{2}S\,2p^{4}(^{3}_{2}P)~^{4}P_{5/2}$                      & 1081480& 1080918   &     1081687  &      4.81E-11 &        4.82E-11  &  2s+1(1)1 2p+2(4)5                         \\
7   & $2s~^{2}S\,2p^{4}(^{3}_{2}P)~^{4}P_{3/2}$                      & 1248820& 1248269   &     1249040  &      2.42E-11 &        2.42E-11  &  2s+1(1)1 2p+2(4)3                         \\
8   & $2s~^{2}S\,2p^{4}(^{3}_{2}P)~^{4}P_{1/2}$                      & 1272280& 1271834   &     1272539  &      2.27E-11 &        2.27E-11  &  2s+1(1)1 2p+2(0)1                         \\
\end{longtable}
\begin{enumerate}[]
\item $^a$ The number at the end or inside of the bracket is 2$J$.
\item $^b$ $s+ = s_{1/2}$, $p- = p_{1/2}$, $p+ = p_{3/2}$, $d- = d_{3/2}$, $d+ = d_{5/2}$, $f- = f_{5/2}$,  and $f+ = f_{7/2}$.
\item $^c$ The number after $\pm$ is the occupation number of the corresponding sub-shell. For example, the $jj$-coupled CSF of level 9 is $2s_{1/2} 2p_{1/2}2p_{3/2}^{3}$.

%\begin{enumerate}[]
%	\item Only the lowest 10 levels are shown here. Table~\ref{tableE} is available in its entirety on the \emph{JQSRT} website.
\end{enumerate}
\clearpage

\scriptsize
\setlength{\tabcolsep}{3pt}
\begin{longtable}{rrcccccccccl}
	\caption{\label{tab:3}Wavelengths ($\lambda$, in vacuum, $\AA$), transition rates ($A$, in s$^{-1}$), oscillator strengths ($gf$, dimensionless) and line strengths ($S$, in a.u.) for the transitions among the 272 levels listed in Table~1. The last column represent the estimated accuracies of the $A$-values using the terminologies of the NIST ASD.}\\
	\toprule
	Upper & Lower & Type & $\lambda_{\rm MCDHF/RCI}$ & $\lambda_{\rm MBPT}$ &  $S_{\rm MCDHF/RCI}$ & $gf_{\rm MCDHF/RCI}$ & $A_{\rm MCDHF/RCI}$&$S_{\rm MBPT}$ & $gf_{\rm MBPT}$ &$A_{\rm MBPT}$ &    Acc.\\
	\midrule
	\endfirsthead
	\caption{(\emph{continued})} \\
	\toprule
	Upper & Lower & Type & $\lambda_{\rm MCDHF/RCI}$ & $\lambda_{\rm MBPT}$ &  $S_{\rm MCDHF/RCI}$ & $gf_{\rm MCDHF/RCI}$ & $A_{\rm MCDHF/RCI}$&$S_{\rm MBPT}$ & $gf_{\rm MBPT}$ &$A_{\rm MBPT}$ &    Acc.\\
	\midrule
	\endhead
	\endfoot
	\bottomrule
	\endlastfoot
2   & 1  & M1&        4.2761E+02&     4.2771E+02&     2.70E+00&  2.55E-05&	2.33E+05&   2.70E+00&	2.55E-05&	2.33E+05 &  B+ \\
2   & 1  & E2&        4.2761E+02&     4.2771E+02&     4.65E-04&  9.99E-10&	9.11E+00&   4.65E-04&	9.97E-10&	9.09E+00 &  B+ \\
3   & 1  & M1&        3.1871E+02&     3.1884E+02&     2.71E-01&  3.44E-06&	3.76E+04&   2.71E-01&	3.44E-06&	3.76E+04 &  B+ \\
3   & 1  & E2&        3.1871E+02&     3.1884E+02&     1.10E-03&  5.71E-09&	6.25E+01&   1.10E-03&	5.70E-09&	6.23E+01 &  B+ \\
4   & 1  & M1&        2.3435E+02&     2.3443E+02&     3.51E-01&  6.06E-06&	3.68E+05&   3.51E-01&	6.06E-06&	3.68E+05 &  B+ \\
4   & 1  & E2&        2.3435E+02&     2.3443E+02&     2.07E-04&  2.70E-09&	1.64E+02&   2.07E-04&	2.70E-09&	1.64E+02 &  B+ \\
4   & 2  & M1&        5.1855E+02&     5.1876E+02&     1.37E-01&  1.07E-06&	1.32E+04&   1.36E-01&	1.06E-06&	1.31E+04 &  B+ \\
4   & 2  & E2&        5.1855E+02&     5.1876E+02&     1.11E-03&  1.34E-09&	1.66E+01&   1.11E-03&	1.34E-09&	1.66E+01 &  B+ \\
\end{longtable}
\begin{enumerate}[]
	\item Only transitions among the lowest 15 levels of the $n=2$ configurations are shown here. Table~\ref{tab:3} is available online in its entirety in the \emph{JQSRT} website.
\end{enumerate}
\clearpage
{\scriptsize
\begin{longtable}{rlccccccccccc}
	\caption{Energies ($E$ in cm$^{-1}$) relative to the ground state for the lowest  272 levels arising from the $2s^{2} 2p^{3}$, $2s 2p^{4}$, $2p^{5}$, $2s^{2} 2p^{2} 3l$~($l=s,p,d$), $2s 2p^{3}3l$ ($l=s,p,d$), and $2p^{4} 3l$ ($l=s,p,d$) configurations of Ge XXVI. AS$_1$,  AS$_2$, AS$_3$, AS$_4$, and AS$_5$ -- the present MCDHF excitation energies; RCI1 and RCI2 -- the present RCI excitation energies, excluding the Breit and QED effects, and 
	including the Breit and QED effects, respectively, using the MCDHF/RCI method; MBPT1 and MBPT2 -- the present MBPT excitation energies, excluding the Breit and QED effects, and 
	including the Breit and QED effects, respectively, using the MBPT method; NIST -- the compiled values from the NIST ASD. \label{tab:1}}\\
	\toprule
	& &  \multicolumn{7}{c}{MCDHF/RCI} &\multicolumn{2}{c}{MBPT} &NIST \\
\cmidrule(r){3-9}\cmidrule(r){10-11}
Key & State &  AS$_1$ & AS$_2$ & AS$_3$ & AS$_4$ & AS$_5$ & RCI1 & RCI2 & MBPT1 & MPBT2 \\
	\midrule
	\endfirsthead
	\caption{(\emph{continued})} \\
	\toprule
	& &  \multicolumn{7}{c}{MCDHF/RCI} &\multicolumn{2}{c}{MBPT} &NIST \\
\cmidrule(r){3-9}\cmidrule(r){10-11}
Key & State &  AS$_1$ & AS$_2$ & AS$_3$ & AS$_4$ & AS$_5$ & RCI1 & RCI2 & MBPT1 & MPBT2 \\\midrule
	\endhead
	\endfoot
	\bottomrule
	\endlastfoot
1&   $2s^{2}\,2p^{3}(^{4}_{3}S)~^{4}S_{3/2}^{\circ}$                &         0&        0&        0&        0&        0&        0&        0&         0&        0&       0\\
2&   $2s^{2}\,2p^{3}(^{2}_{3}D)~^{2}D_{3/2}^{\circ}$                &    236799&   236226&   236151&   236106&   236094&   236188&   233859&    236131&   233805&  233740\\
3&   $2s^{2}\,2p^{3}(^{2}_{3}D)~^{2}D_{5/2}^{\circ}$                &    324198&   322932&   322756&   322655&   322624&   322745&   313767&    322607&   313636&  313520\\
4&   $2s^{2}\,2p^{3}(^{2}_{1}P)~^{2}P_{1/2}^{\circ}$                &    431628&   429819&   429403&   429212&   429129&   429434&   426704&    429316&   426572&  426510\\
5&   $2s^{2}\,2p^{3}(^{2}_{1}P)~^{2}P_{3/2}^{\circ}$                &    640905&   639754&   639492&   639365&   639315&   639695&   630081&    639638&   630053&  629730\\
6&   $2s~^{2}S\,2p^{4}(^{3}_{2}P)~^{4}P_{5/2}$                      &   1093017&  1090117&  1090295&  1090428&  1090495&  1089307&  1080918&   1089667&  1081687& 1081480\\
7&   $2s~^{2}S\,2p^{4}(^{3}_{2}P)~^{4}P_{3/2}$                      &   1260953&  1257723&  1257881&  1258013&  1258079&  1256897&  1248269&   1257246&  1249040& 1248820\\
8&   $2s~^{2}S\,2p^{4}(^{3}_{2}P)~^{4}P_{1/2}$                      &   1283134&  1279555&  1279597&  1279693&  1279741&  1278625&  1271834&   1278919&  1272539& 1272280\\
\end{longtable}
}

\clearpage

\begin{longtable}{clcccc}
\caption{\label{tab:5}The total energies $E_{h}$ (a.u.), hyperfine magnetic dipole constants $A_{J}$($I$/$\mu$$_{I}$) (MHz per unit of $\mu$$_N$ ), electric quadrupole constants $B_{J}/Q$ (MHz/barn), and Land\'e $g_{J}$-factors from the present MCDHF/RCI calculations. }\\
	\toprule
Key &State & $E$(a.u.) & $A$(MHz) & $B$(MHz) & 
$g_J$ \\
	\midrule
	\endfirsthead
	\caption{(\emph{continued})} \\
	\toprule
Key &State & $E$(a.u.) & $A$(MHz) & $B$(MHz) & 
$g_J$ \\	\midrule
	\endhead
	\endfoot
	\bottomrule
	\endlastfoot
1   & $2s^{2}\,2p^{3}(^{4}_{3}S)~^{4}S_{3/2}^{\circ}$                &              -1548.763283  &           3.066E+04 &	     4.573E+04 &	   1.680406E+00 \\
2   & $2s^{2}\,2p^{3}(^{2}_{3}D)~^{2}D_{3/2}^{\circ}$                &              -1547.697743  &           2.480E+04 &	     6.897E+04 &	   1.204494E+00 \\
3   & $2s^{2}\,2p^{3}(^{2}_{3}D)~^{2}D_{5/2}^{\circ}$                &              -1547.333653  &           1.037E+05 &	    -5.956E+02 &	   1.194149E+00 \\
4   & $2s^{2}\,2p^{3}(^{2}_{1}P)~^{2}P_{1/2}^{\circ}$                &              -1546.819074  &           3.006E+05 &	     0.000E+00 &	   6.583320E-01 \\
5   & $2s^{2}\,2p^{3}(^{2}_{1}P)~^{2}P_{3/2}^{\circ}$                &              -1545.892421  &           5.860E+04 &	    -1.157E+05 &	   1.233373E+00 \\
6   & $2s~^{2}S\,2p^{4}(^{3}_{2}P)~^{4}P_{5/2}$                      &              -1543.838256  &           2.368E+05 &	     8.035E+04 &	   1.563662E+00 \\
7   & $2s~^{2}S\,2p^{4}(^{3}_{2}P)~^{4}P_{3/2}$                      &              -1543.075749  &           1.274E+05 &	    -7.865E+04 &	   1.647208E+00 \\
8   & $2s~^{2}S\,2p^{4}(^{3}_{2}P)~^{4}P_{1/2}$                      &              -1542.968381  &           6.258E+05 &	     0.000E+00 &	   2.535795E+00 \\
\end{longtable}

\end{document}